\documentclass[10pt,aps,twocolumn,nofootinbib]{revtex4}
\usepackage{amsmath,graphics,epsfig}
\usepackage{color,hyperref}
\usepackage{datetime}
\begin{document}

\title{Towards Realistic SUSY Grand Unification for Extended MSSM }
\author{Sibo Zheng}
\email{sibozheng.zju@gmail.com}
\affiliation{\small{Department of Physics, Chongqing University, Chongqing 401331, P. R. China}}
\date{March 27, 2019}
\begin{abstract}
Low-energy supersymmetric models such as MSSM, NMSSM and MSSM with vectorlike fermion are consistent with perturbative unification.
While the non-minimal extensions naturally explain Higgs mass and dark matter in the low energy region,
it is unclear how seriously they are constrained in the ultraviolet region.
Our study shows that $i)$, In the case of embedding MSSM into $\rm{SU}(5)$, the fit to SM fermion masses requires a singlet $S$,
which leads to unviable embedding of NMSSM into $\rm{SU}(5)$ because such $S$ feeds singlet $N$ a mass of order unification scale as well.
$ii)$, Similar result holds in the case of embedding NMSSM into $\rm{SO}(10)$, 
where $S$ is replaced by some Higgs fields responsible for $\rm{SO}(10)$ breaking.
$iii)$, On the contrary, for the embedding of MSSM with $16$-dimensional vectorlike fermions into $\rm{SO}(10)$,
the Higgs field responsible for the vectorlike mass of order TeV scale 
can evade those problems the singlet $N$ encounters because of an intermediate mass scale in the $126$-dimensional Higgs field.
\end{abstract}
\maketitle

\section{Introduction}
At the frontiers of new physics beyond standard model (SM) 
natural or TeV-scale supersymmetry (SUSY) offers us a grand unification (GUT) of SM gauge coupling constants \cite{unif1,unif2,unif3,unif4}.
Such natural SUSY hosts a lot of SUSY particles which can be directly detected at the particle collider LHC 
or dark matter direct detection facilities such as Xenon-1T.
Meanwhile, embedding these TeV-scale SUSY models into the ultraviolet completions - SUSY GUT- 
may solve the long-standing issues such as SM flavor puzzle and neutrino masses.

Nowadays experimental data seems to oppose the minimal supersymmetric standard model (MSSM)  
either from the bottom or top viewpoint.
In the former one, 
the $125$ GeV Higgs mass \cite{Higgsmass1, Higgsmass2} requires either large mixing effect or soft masses of order $10$ TeV for the stop scalars \cite{MSSMHiggs1, MSSMHiggs2, MSSMHiggs3}. 
When the mixing effects among generations are significant,
the constraints from flavor violation tend to require the SUSY mass order far above the weak scale.
Moreover, the direct detection limits of dark matter \cite{MSSMDark1,MSSMDark2} impose rather strong pressure on the scenario of neutralino dark matter.
In the later perspective,
the minimal SUSY $\rm{SU}(5)$ referring to MSSM is significantly constrained by the proton decay \cite{protonreivew}. 
It requires the color-triplet Higgs mass of order GUT scale, which together with unification, 
leads to the MSSM mass spectrum at least of order $100$ TeV \cite{Murayama, Hisano,Bajc}.

Therefore, it is of great interest to explore the MSSM with rational extensions that can resurrect the natural SUSY once again. 
Along this direction, there are at least two simple examples -the next-to-minimal supersymmetric standard model (NMSSM) \cite{NMSSM} 
and the MSSM with vectorlike (VL) fermions (VMSSM) \cite{0807.3055} -
which are consistent with unification \cite{Zheng1, Zheng2}.
While these extensions provide natural explanations of Higgs mass and dark matter in the low energy region,
it is unclear what the statuses of them are in the ultraviolet energy region.
This is the main focus of this study.

In this paper, 
we discuss the embeddings of MSSM, NMSSM and VMSSM into realistic grand unification (GUT) \cite{GG, SO101, SO102}.
In each case, both the SUSY $\rm{SU}(5)$ \cite{SUSYSU51,SUSYSU52, SUSYSU53} and SUSY $\rm{SO}(10)$ \cite{SUSYSO10} representations will be explored.
In these SUSY GUTs, we discuss the GUT-scale superpotential 
\begin{eqnarray}{\label{W}}
W=W_{Y}+W_{\text{SB}},
\end{eqnarray}
with following features:
\begin{itemize}
\item $W_Y$ and $W_{\text{SB}}$ are both renormalizable.
\item All of vacuum expectation values (vevs) are dynamically generated from $W_{\text{SB}}$.
\item All of SM matters and extended matters obtain their masses via the Higgs mechanism in $W_Y$. 
\end{itemize}

Since $W_Y$ is fixed by the SM and the extra matters such as the singlet $N$ or VL fermions at TeV scale,
 it is crucial to find suitable content of $W_{\text{SB}}$ that achieves the breaking of gauge group 
$G_{\text{GUT}}\rightarrow G_{\text{SM}}$.
In Sec.II, we explore the embedding of MSSM into realistic SUSY GUTs,
where useful conventions and notation will be introduced.
The analysis on the MSSM is of great use to guide us towards the embeddings of NMSSM and VMSSM.
Sec.III and Sec.IV is devoted to study the embeddings of NMSSM and VMSSM into realistic SUSY GUTs, respectively. 
Finally, we conclude in Sec.V.

\section{The Benchmark Model: MSSM}
In the minimal $\rm{SU}(5)$ of Standard Model (SM) gauge group, 
the SM fermions of each generation are assigned as $1$, $\bar{5}$ ($\psi$) and $10$ ($\Phi$) of $\rm{SU}(5)$ for right-hand neutrino $N_{R}$, 
$L$ and down quark $\bar{d}$, and $Q$, up quark $\bar{u}$ and $\bar{e}$, respectively;
whereas in the $\rm{SO}(10)$ representation, 
the SM fermions of each generation are embedded into a $16$-dimensional representation, 
which decomposes as $16=1+\bar{5}+10$ under the $\rm{SU}(5)$.

\subsection{\rm{SU}(5)}
The Yukawa superpotenail $W_Y$ in Eq.(\ref{W}) contains two parts
\begin{eqnarray}{\label{Yukawa5}}
W^{\rm{SU}(5)}_{Y}=W^{\rm{SU}(5)}_{Y_{f}}+W^{\rm{SU}(5)}_{Y_{\nu}}
\end{eqnarray}
which refers to SM fermions without neutrinos and neutrinos, respectively.

According to the product $\bar{5}\times 10=5+45$ and $10\times10=\bar{5}_{s}+\bar{45}_{a}+\bar{50}_{s}$, 
where subscript ``s'' and ``a" refers to symmetric and anti-symmetric, respectively.
With the Higgs representations composed of $5$, $\bar{5}$ and $\bar{45}$ \cite{GJ},
$W^{\rm{SU}(5)}_{Y_{f}}$ is given by,
\begin{eqnarray}{\label{Yukawa5f}}
W^{\rm{SU}(5)}_{Y_{f}}&=&Y_{d}^{ij}\psi_{ai}(\bar{5})\bar{H}_{b}(\bar{5})\Phi^{ab}_{j}(10)\nonumber\\
&+&Y_{45}^{ij}\psi_{ai}(\bar{5})\bar{H}^{a}_{bc}(\bar{45})\Phi^{bc}_{j}(10)\nonumber\\
&+&Y_{u}^{ij}\epsilon_{abcde}\Phi^{ab}_{i}(10)H^{e}(5)\Phi^{cd}_{j}(10)
\end{eqnarray}
where $a, b, c$, etc denote the $\rm{SU}(5)$ indexes, $i, j$ are the generation indexes,
and $Y_{u,d, 45}$ are Yukawa matrixes.
Note, $\Phi^{ab}$ is an anti-symmetric field and $H^{a}_{bc}=-H^{a}_{cb}$.

The reason to include $\bar{45}$ is clear in the SM fermion mass matrixes as derived from Eq.(\ref{Yukawa5f}):
\begin{eqnarray}{\label{massmatrix1}}
M_{u}&=&Y_{u}^{ij}\upsilon^{5}_{u}, \nonumber\\
M_{d}&=&Y_{d}^{ij}\upsilon^{\bar{5}}_{d}+Y^{ij}_{45}\upsilon^{\bar{45}}_{d}, \nonumber\\
M_{e}&=&Y_{d}^{ij}\upsilon^{\bar{5}}_{d}-3Y^{ij}_{45}\upsilon^{\bar{45}}_{d}.
\end{eqnarray}
where $\upsilon^{5}_{u}$ and $\upsilon^{\bar{5}}_{d}$ is vev of Higgs doublet $H_{u}$ and $H_{d}$ in $5$ and $\bar{5}$, respectively;
$\upsilon_{45}$ is the vev of doublet $\sigma_{d}$ in $\bar{45}$,  
which is defined as $\left<H^{b5}_{a}\right>=\upsilon_{45}(1,1,1,-3)_{\text{diag}}$ for $a,b=1-4$.
Without the $\bar{45}$,  $M_{d}=M_{e}$ at the GUT scale for all of three generations. 
While such mass relation is viable for the third generation, 
they should be avoided for the first two generations.
Adding $\bar{45}$ can tune the incorrect mass relations to the desired ones,
\begin{eqnarray}{\label{app}}
m_{d}\simeq 3 m_{e},~m_{s}\simeq m_{\mu}/3,~m_{b}\simeq m_{\tau},
\end{eqnarray}

The GUT-scale mass relations in Eq.(\ref{app}) strongly constrain the Yukawa matrixes $Y_{u,d, 45}$.
For example, some specific choices on $Y_{u,d, 45}$ in the Georgi-Jarlskog scheme \cite{GJ} lead to a stable $b$ quark. 
In order to solve the SM flavor issue, we choose the Fritzsch scheme \cite{Fritzsch},
in which $Y_{u,d, 45}$ take the following forms
\begin{eqnarray}{\label{Fritzsch}}
Y_{u}^{ij}\upsilon_{u}=
\left(%
\begin{array}{ccc}
0 & A_{u} &0 \\
A_{u} & 0 & B_{u} \\
0 & B_{u} & C_{u}\\
\end{array}%
\right),\nonumber\\
Y_{d}^{ij}\upsilon_{d}=
\left(%
\begin{array}{ccc}
0 & A_{d} & 0 \\
A_{d} & 0 & B_{d} \\
0 & B_{d} & C_{d}\\
\end{array}%
\right),\nonumber \\
Y_{45}^{ij}\upsilon_{45}=
\left(%
\begin{array}{ccc}
0 &  0 & 0 \\
0 & 0 & D_{d} \\
0 & D_{d} & 0 \\
\end{array}%
\right),
\end{eqnarray}
where there are small mass hierarchies $C_{f}>>B_{f}>>A_{f}$ with $f=\{u,d,e\}$ in Eq.(\ref{Fritzsch}) so as to address the SM flavor mass hierarchies.
Substituting Eq.(\ref{Fritzsch}) into Eq.(\ref{app}) implies that there is a fine tuning between $B_{d}$ and $D_{d}$,
\begin{eqnarray}{\label{tuning}}
D_{d}=(1\pm \frac{2\sqrt{3}}{3})B_{d}.
\end{eqnarray}
With this fine-tuning solution,  the diagonalizations of matrixes in Eq.(\ref{Fritzsch}) yield the CKM matrix
\begin{eqnarray}{\label{s}}
\mid V_{\text{CKM}}\mid \simeq
\left(%
\begin{array}{ccc}
0.974 & 0.227 & 0.004 \\
0.227 & 0.970 & 0.042 \\
0.008 & 0.042 & 0.999 \\
\end{array}%
\right),
\end{eqnarray}
which is in good agreement with experimental data.

For $W_{Y_{\nu}}$ responsible neutrino masses we take a simple form as follows,
\begin{eqnarray}{\label{Yukawan}}
W^{\rm{SU}(5)}_{Y_{\nu}}&=&Y_{N}^{ij}N_{Ri}(1)\psi_{aj}(\bar{5})H^{a}(5)\nonumber\\
&+& Y_{S}^{ij}S(1) \cdot N_{Ri}(1) N_{Rj}(1)+\text{H.c}
\end{eqnarray}
where $S$ is a singlet of $G_{\text{SM}}=SU(3)_{c}\times SU(2)_{L}\times U(1)_{Y}$, with $\langle S\rangle$ of order GUT scale.
In Eq.(\ref{Yukawan}), 
one finds the neutrino Dirac mass $M_{\nu}=Y^{ij}_{N}\upsilon^{5}_{u}$ and  
the right-hand neutrino mass $M_{N_{R}}=Y^{ij}_{S}\left<S\right>$,
which results in the left-hand neutrino masses in terms of type-I \cite{typeI1,typeI2,typeI3,typeI4} seesaw mechanism,
\begin{eqnarray}{\label{neutrino}}
m_{\nu}=M^{T}_{\nu}M_{N_{R}}^{-1}M_{\nu}=Y_{N}^{T}Y^{-1}_{S}Y_{N}\frac{\upsilon^{5}_{u}}{\langle S\rangle}.
\end{eqnarray}
Given $\upsilon^{5}_{u}\sim 10^2$ GeV, $\langle S\rangle \sim 10^{15}-10^{16}$ GeV and Yukawa couplings of order unity,
the neutrino mass is of order $\sim 10^{-2}-10^{-3}$ eV.
Similar to $Y_{u,d,e}$,  $Y_{S}$ and $Y_{N}$ in Eq.(\ref{Yukawan}) are also constrained 
by the fit to neutrino mixings as described by PMNS matrix $U_{\text{PMNS}}=U_{e}^{-1}U_{\nu}^{-1}$,
where $U_{\nu}$ and $U_{e}$ are defined to diagonalize mass matrix $m_{\nu}$ and $M_e$, respectively.

After we have established on a benchmark solution\footnote{It is of special interest to examine whether there is viable solution to the input parameters in the case that all mass matrixes such as $M_{u,d,e, N_{R}}$ are assigned the Fritzsch form.}
to the input parameters at the GUT scale which can explain the SM flavor issue and neutrino masses,
we turn to the structure of $W_{\text{SB}}$.
$1)$. In order to obtain light neutrino masses, $\langle S\rangle$ of order GUT scale is required.
$2)$. In order to break the $\rm{SU}(5)$ we introduce $75(Z)$.
With such $75$, we can add a $50$ and $\bar{50}$ to achieve the doublet-triplet splitting \cite{0007254} for $5$ and $\bar{5}$.
$3)$. In order to gain nonzero vev $\upsilon^{\bar{45}}_{d}$, we include another $75(Z')$ with a vev of the GUT scale.
The reason for this is that neither $1$ nor $24$ with large vev is favored by the product $H(5)\times \bar{H}(\bar{5})=1+24$.
$4)$, Due to the singlet $S$ there is an unsafe operator
\begin{eqnarray}{\label{unsafe5}}
W_{\text{unsafe}}=SH(5)\bar{H}(\bar{5}),
\end{eqnarray}
which must be eliminated.

Shown in Table.\ref{MSSMc} is the $Z_{2}\times Z'_{2}$ parity, 
which can eliminate the unsafe operator in Eq.(\ref{unsafe5}).
Under this parity, $W_{\text{SB}}$ reads as,
\begin{eqnarray}{\label{SSBM5}}
W^{\rm{SU}(5)}_{\text{SB}}&=& \frac{M_{Z}}{2} Z^{2} +\frac{M_{Z'}}{2} Z'^{2}+\frac{M_{S}}{2}S^{2}+\frac{M_{S'}}{2}S'^{2}\nonumber\\
&+&\frac{\lambda_{Z}}{3} Z^{3}+\frac{\lambda_{S}}{3}S^{3}\nonumber\\
&+&\lambda_{1}SZ^{2}+\lambda_{2}SZ'^{2}+ \lambda_{3}ZZ'^{2}+\lambda_{4}SS'^{2}\nonumber\\
&+&\bar{H}(\bar{45}) \left(M_{45}+\lambda_{5}S\right)H(45) \nonumber\\
&+&\lambda_{6}S' \bar{H}(\bar{50})H(50) \nonumber\\
&+& \lambda_{7} Z \bar{H}(\bar{5})H(50)+\lambda_{8}Z' H(5)\bar{H}(\bar{50})\nonumber\\
&+& \lambda_{9}Z H(45)\bar{H}(\bar{5})+\text{H.c}.
\end{eqnarray}
According to the $F$ terms in Eq.(\ref{SSBM5}), 
the nonzero singlet vevs $ \langle(1,1,1)_{Z}\rangle=Z$, $\langle(1,1,1)_{Z'}\rangle=Z'$, $\langle(1,1,1)_{S}\rangle=S$ and $\langle(1,1,1)_{S'}\rangle=S'$ are given by,
\begin{eqnarray}{\label{vevs}}
\frac{S}{M_{Z}}&=&-\frac{1}{2\lambda_{4}}\eta_{3}= -\frac{1}{2\lambda_{2}}\left(-2\lambda_{3}\frac{Z}{M_{Z}}+\eta_{1}\right), \nonumber\\
\frac{Z'}{M_{Z}}&=&\frac{1}{\sqrt{\lambda_{S}}}\left(\frac{-Z}{M_{Z}}\right)^{\frac{1}{2}}\left[1+\lambda_{Z}+\frac{\lambda_{1}}{\lambda_{2}}\left(-2\lambda_{3}\frac{Z}{M_{Z}}+\eta_{1}\right)\right]^{\frac{1}{2}} \nonumber\\
\frac{Z}{M_{Z}}&=&\frac{-b\pm \sqrt{b^{2}-4ac}}{2a},
\end{eqnarray}
where 
\begin{eqnarray}{\label{abc}}
a&=&\lambda_{1}+\frac{\lambda^{2}_{3}}{\lambda^{2}_{2}}\lambda_{S}-\frac{\lambda_{2}\lambda_{Z}}{\lambda_{S}}+2\frac{\lambda_{1}\lambda_{3}}{\lambda_{S}}\nonumber\\
b&=&-\frac{\lambda_{3}}{\lambda_{2}}-\frac{\lambda_{3}\lambda_{S}}{\lambda^{2}_{2}}\eta-\frac{\lambda_{2}}{\lambda_{S}}-\frac{\lambda_{1}}{\lambda_{S}}\eta\nonumber\\
c&=&\frac{1}{2\lambda_{2}}\eta_{1}\eta_{2}+\frac{\lambda_{S}}{4\lambda^{2}_{2}}\eta_{1}^{2}+\lambda_{4}\left(\frac{S'}{M_{Z}}\right)^{2}
\end{eqnarray}
and $\eta_{1}=M_{Z'}/M_{Z}$, $\eta_{2}=M_{S}/M_{Z}$, $\eta_{3}=M_{S'}/M_{Z}$.

A few comments are in order regarding the parity assignments. 
Firstly, the $Z_2$ parity eliminates the unsafe operator in Eq.(\ref{unsafe5}).
Secondly, without $Z'_2$ large $\mu$ term for the doublets in $5$ and $\bar{5}$ 
would be induced by the mixings with $45$ and $\bar{45}$.
Instead, imposing $Z'_2$ forbids the operator $Z' H(5)\bar{H}(\bar{45})$, 
which then keeps the doublets in $5$ and $\bar{5}$ light.
Finally, due to the last line in Eq.(\ref{SSBM5}) which is consistent with $Z_{2}\times Z'_{2}$,
the effective operator $\bar{H}(\bar{5})H(45)Z'^{2}/M_{Z}$ is produced after integrating $Z$.
Thus, the effective superpotential for the doublets in $45$ and $\bar{45}$ at the leading order is given by,
\begin{eqnarray}{\label{eff1}}
W_{\text{eff}}\sim \left(M_{45}+\lambda_{5}\langle S\rangle\right)\sigma_{u}\sigma_{d}+ \lambda_{3}H_{d}\sigma_{u}Z'^{2}/M_{Z},
\end{eqnarray}
where corrections due to those mixings among singlets of $Z$, $Z'$ and $S$ have been neglected. 
From Eq.(\ref{eff1}) we obtain the vev 
\begin{eqnarray}{\label{v45}}
\sigma_{d}\sim \lambda_{3}\frac{Z'^{2}}{(M_{45}+\lambda_{4}\langle S\rangle)M_{Z}}\upsilon_{d}.
\end{eqnarray}
Given $\lambda_{3}\sim 0.1$, $\upsilon_{d} \sim 10$ GeV,
we have $\sigma_{d}\sim 100$ MeV. 
A rational Yukawa texture such as $Y_{45}\sim (0.01,0.1, 1)$ for the three generations then reproduces the mass relations in Eq.(\ref{app}).

\begin{table}
\tiny{
\begin{center}
\begin{tabular}{|c|c|c|c||c|c|c||c|c|c|c|c|c|c|}
\hline
 Field & \ $N_{R}$&\ $\psi$ & \ $\Phi$   & \  $5$ & \ $\bar{5}$ & \ $\bar{45}$ & \ $1(S)$& \ $1(S')$& \ $75(Z)$ &\ $75(Z')$ & $45$ & \ $50$ &  $\bar{50}$   \\
  \hline
$Z_{2}$ &+ & + &  - & + & - & -  & + & + & + & - & - & -  & -  \\
$Z'_{2}$ &+ & + &  + & + & + & + & + & - & + & - & + & +  & -  \\
  \hline
\end{tabular}
\caption{$Z_{2}\times Z'_{2}$ parity assignments in the case of embedding MSSM into $\rm{SU}(5)$,
which are consistent with the superpotentials in Eq.(\ref{Yukawa5}) and Eq.(\ref{SSBM5}). }
\label{MSSMc}
\end{center}}
\end{table}

\subsection{\rm{SO}(10)}
Unlike the case of $\rm{SU(5)}$, 
the input parameters in the $\rm{SO(10)}$ which control the SM fermions masses $M_{u,d,e}$ 
and neutrino masses $m_{\nu}$ are tied to each other.
The main reason for this is that the MSSM matters of each generation are contained in a single $16(\phi)$.
Here, we give a brief review on the embedding of MSSM into $\rm{SO}(10)$.

According to the product $16\times 16= 10_{s}+120_{a}+126_{s}$, 
the Higgs representation responsible for SM fermion masses can be composed of $10$, $120$ and $\bar{126}$.
By following previous discussions on $\rm{SU}(5)$ \cite{GN1, GN2}, the simplest choice is to introduce $10$ and $\bar{126}$ \cite{9209215}, with Yukawa superpotential given by,
\begin{eqnarray}{\label{Yukawa10}}
W^{\rm{SO}(10)}_{Y}=\phi_{i}(16)[Y^{ij}_{10}H(10)+Y^{ij}_{126}\bar{H}(\bar{126})]\phi_{j}(16),
\end{eqnarray}
where matrixes $Y_{10}$ and $Y_{126}$ are both symmetric.
From Eq.(\ref{Yukawa10}) we have
\begin{eqnarray}{\label{massmatrix2}}
M_{u}&=&Y_{10}^{ij}\upsilon^{10}_{u}+Y_{126}^{ij}\upsilon^{\bar{126}}_{u}, \nonumber\\
M_{d}&=&Y_{10}^{ij}\upsilon^{10}_{d}+Y_{126}^{ij}\upsilon^{\bar{126}}_{d}, \nonumber\\
M_{e}&=&Y_{10}^{ij}\upsilon^{10}_{d}-3Y_{126}^{ij}\upsilon^{\bar{126}}_{d},
\end{eqnarray}
for SM quarks and electrons, and 
\begin{eqnarray}{\label{massmatrix3}}
M_{N_{R}}&=&Y_{126}^{ij}\upsilon^{\bar{126}}_{s} \nonumber\\
M_{\nu}&=&Y_{10}^{ij}\upsilon^{10}_{u}-3Y_{126}^{ij}\upsilon^{\bar{126}}_{d},\nonumber\\
M_{L}&=&Y_{126}^{ij}\upsilon^{\bar{126}}_{L}
\end{eqnarray}
for neutrinos,
where $\upsilon^{10}_{u,d}$ refer to the doublet vevs in $H(10)$, 
and $\upsilon^{\bar{126}}_{s, u, d, L}$ denote the singelt (s), doublet (u, d) and triplet (L) vevs in $\bar{H}(\bar{126})$, respectively.
Therefore, the neutrino mass arises from both type-I \cite{typeI1,typeI2,typeI3,typeI4} and type-II \cite{typeII1,typeII2,typeII3} contributions,
\begin{eqnarray}{\label{seesaw}}
m_{\nu}=M_{L}-M^{T}_{\nu}M_{N_{R}}^{-1}M_{\nu}.
\end{eqnarray}

Benchmark solutions to the input parameters in Eq.(\ref{massmatrix2}) and Eq.(\ref{massmatrix3}) 
have been extensively studied \cite{9209215, SO10fit1, SO10fit2, SO10fit3, SO10fit4},
which demonstrate that the fit to SM flavor masses and mixings is viable.
But the construction of $W_{\text{SB}}$ responsible for the breaking of $\rm{SO}(10)\rightarrow G_{\text{SM}}$ is challenging \cite{SO10Mp,SO10Mp1,SO10Mp2, SO10Mp3},
since some of triplet fields in $126$ and $\bar{126}$ obtain mass of order $\upsilon^{\bar{126}}_{s}\sim 10^{13}$ GeV,
which spoils the perturbative unification. 
Attempts to solve this issue involve adding $54$ \cite{1805.10631} to $W_{\text{SB}}$
or adding $120$ \cite{1805.05776} to $W_{Y}$ in Eq.(\ref{Yukawa10}).

We employ the solution of modifying $W_{\text{SB}}$ \cite{1805.10631},
where the Higgs fields are composed of $210(Y)$, $H(126)$, $\bar{H}(\bar{126})$ and $54(X)$,
and $W_{\text{SB}}$ takes the form\footnote{We follow the notation in \cite{0405300} for Yukawa coupling constants.}
\begin{eqnarray}{\label{SSBM10}}
W^{\text{SO}(10)}_{\text{SB}}&=& \frac{M_{Y}}{2} Y^{2} +\frac{M_{X}}{2}X^{2}+ M_{126} H(126)\bar{H}(\bar{126})\nonumber\\
&+&\lambda_{1} Y^{3}+\lambda_{2} Y H(126)\bar{H}(\bar{126}) \nonumber\\
&+& Y\left[\lambda_{3}H(126)H(10)+\lambda_{4}\bar{H}(\bar{126})H(10)\right]\nonumber\\
&+&X\left[\lambda_{8}X^{2}+\lambda_{10}Y^{2}+\lambda_{11}H^{2}(126)\right.\ \nonumber\\
&+&\left.\  \lambda_{12}\bar{H}^{2}(\bar{126})+ \lambda_{13}H^{2}(10)\right]+ \text{H.c}.
\end{eqnarray}

Here are a few comments about $W_{\text{SB}}$ in order.
$1)$. Under the notation of $SU(4)\times SU(2)_{L}\times SU(2)_{R}$ 
the SM singlet vevs $Y_{1}=\langle(1,1,1)_{Y}\rangle$, $Y_{2}=\langle(15,1,1)_{Y}\rangle$ and
$Y_{3}=\langle(15,1,3)_{Y}\rangle$ in  $Y$ and $X=\langle(1,1,1)_{X}\rangle$ in $X$ 
are responsible for the breaking $\rm{SO}(10)\rightarrow G_{\text{SM}}\times U(1)_{B-L}$.
$2)$. The SM singlet vevs $\upsilon^{126}_{s}=\langle(\overline{10},1,3)_{126}\rangle$ and 
$\upsilon^{\bar{126}}_{s}=\langle(10,1,3)_{\bar{126}}\rangle$ 
result in the breaking of $G_{\text{SM}}\times U(1)_{B-L}\rightarrow G_{\text{SM}}$.

In the limit $\upsilon^{126}_{s}=\upsilon^{\bar{126}}_{s}<<Y_{i}, X$ the vevs are given by \cite{1805.10631},
\begin{eqnarray}{\label{vevs3}}
\frac{Y_1}{Y_2}&=&\frac{1}{2\sqrt{3}}\eta^2,\nonumber\\
\frac{M_Y}{Y_2}&=&-\frac{\lambda_1}{5\sqrt{2}}\left(3+\eta^2\right),\nonumber\\
\frac{X}{Y_2}&=&\frac{\lambda_1}{\lambda_{10}\sqrt{30}}\left(-2+\eta^2\right),\nonumber\\
\frac{M_{126}}{Y_2}&=&-\frac{\lambda_2}{120}\left(6\sqrt{2}+12\eta+\sqrt{2}\eta^2\right),\nonumber\\
\frac{M_X}{Y_2}&=&\frac{-1}{20\sqrt{2}\lambda_1\lambda_{10}}\left[\left(5\lambda_{10}^3+2\lambda_{1}^{2}\lambda_{8} \right)
\eta^2+20\lambda_{10}^3-4\lambda_{1}^{2}\lambda_{8}\right]\nonumber\\
\end{eqnarray}
with $\eta=Y_{3}/Y_2$.

\section{NMSSM}
With the embedding of MSSM into realistic SUSY GUTs as a benchmark,
in this section we analyze the NMSSM.
According to the starting points in the Introduction, 
a viable embedding should satisfy two constraints:
\begin{itemize}
\item The mass of $N$ should be of order TeV scale.
\item The vev of $N$ should be of order TeV scale.
\end{itemize}
Both of them may be spoiled by a few dangerous mixings between $N$ and Higgs fields which contain singlet vev of order GUT scale. 
The key point is whether there is suitable symmetry to avoid such mixings.

\subsection{\rm{SU}(5)}
In this situation,  $W_Y$ in Eq.(\ref{Yukawa5}) should be extended to include $NH_{u}H_{d}+\frac{\kappa}{3}N^{3}+\text{H.c}$ in the NNSSM, 
which means that\footnote{Operator $NH(5)\bar{H}(\bar{5})$ contributes to Yukawa interaction $NH_{c}\bar{H}_{c}$ beyond MSSM, 
with $H_c$ and $\bar{H}_c$ being the color-triplet Higgs fields.  
However, it does not affect proton decay at all, 
as the singlet $N$ mass is always far larger than proton mass.},
\begin{eqnarray}{\label{NMSSM5}}
\delta W_{Y}=\lambda NH(5)\bar{H}(\bar{5})+\frac{\kappa}{3}N^{3}.
\end{eqnarray}
Eq.(\ref{NMSSM5}) do not affect the fit to SM flavor masses and mixings in Sec.IIA.
Nevertheless, compared to MSSM, $W_{\text{SB}}$ is allowed to contain superpotential terms
\begin{eqnarray}{\label{SSBNM5}}
 W_{\text{unsafe}}= N Z^{2}+NZ'^{2}+NS^{2}+N^{2}S.
\end{eqnarray}
These new terms in Eq.(\ref{SSBNM5}) yield corrections to the $F$-terms of $Z$, $Z'$, $S$ 
such as $F_{S}=F^{\text{MSSM}}_{S}+2\lambda_{N}SN$, 
which can be adjusted to the case of MSSM by e.g. $\left<N\right>=0$.
Even so, the singlet vevs $\left<Z\right>$, $\left<Z'\right>$ and $\left<S\right>$ still lead to either large $N$ mass or large mixing.

In order to avoid all of mixing terms in Eq.(\ref{SSBNM5}), we need to impose new parity.
The first observation is that an odd $N$ under a $Z_2$ parity as shown in Table.\ref{MSSMc} excludes the first three terms in Eq.(\ref{SSBNM5}).
But the last term therein still remains \footnote{A economic solution to keep light $N$  
is adding another singlet $S'=1$ which is even under the $Z_2$.
With such $S'$ the $W_{\text{SB}}$ is further extended by,
\begin{eqnarray}{\label{S'}}
\delta W_{\text{SB}}(S')=\frac{1}{2}M_{S'}S'^{2}+m_{SS'}SS'+N^{2}S'.
\end{eqnarray}
from which $F_{S'}=M_{S'}S'+m_{SS'}S+N^{2}$,
and $\langle S'\rangle=-\langle S\rangle$ if $M_{S'}=m_{SS'}$.
The two different contributions to $N$ mass cancel each other, leaving us a light $N$. 
Unfortunately, neither Abelian or $Z_N$ parity can ensure $M_{S'}=m_{SS'}$.}. 
Similar result holds for $Z_N$ or an Abelian symmetry.

\subsection{\rm{SO}(10)}
Similar to the embedding of NMSSM into $\rm{SU}(5)$, 
$W_Y$ in Eq.(\ref{Yukawa10}) is modified by,  
\begin{eqnarray}{\label{NMSSM52}}
\delta W_{Y}=\lambda NH^{2}(10)+\frac{\kappa}{3}N^{3}.
\end{eqnarray}
Instead of Eq.(\ref{SSBNM5}), $W_\text{SB}$ in Eq.(\ref{SSBM10}) is allowed by gauge invariance to contain
\begin{eqnarray}{\label{NMSSM102}}
W_{\text{unsafe}}&=&N Y^{2}+NH(126)\bar{H}(\bar{126})+N X^{2}+XH^{2}(10),\nonumber\\
\end{eqnarray}
in which $N$ mixes with the SM singlets of $Y$, $126$, $\bar{126}$ and $X$.
Thus, all of Yukawa couplings in Eq.(\ref{NMSSM102}) have to be extremely small.

What kind of parity allows Yukawa superpotential in Eq.(\ref{NMSSM52}) 
but eliminates that in Eq.(\ref{NMSSM102}) simultaneously?
The first observation is that a $Z_2$ parity does not work.
Since Eq.(\ref{NMSSM102}) would imply $N$ an odd field,
which contradicts with the Yukawa superpotential in Eq.(\ref{NMSSM52}).
Similar results hold for any $Z_N$ parity.
Because the rational assignment $n_{Y}=0$ as required by successful symmetry breaking, 
demonstrates that in order to allow the Yukawa superpotential in  Eq.(\ref{NMSSM52}), $n_{10}=n_{\bar{126}}=n_{126}=N/2$.
Accordingly, $n_{N}=0$ from $N10(H)10(H)$, which implies that some of terms in Eq.(\ref{NMSSM102}) are still allowed.
To conclude, in our setup embedding NMSSM into the minimal $\rm{SO}(10)$ is not viable.\\

\section{VMSSM}
Let us proceed to discuss the embedding of VMSSM into SUSY $\rm{SU(5)}$ and $\rm{SO}(10)$.
The VL fermions with mass of order TeV scale can be composed of $5$ with $\bar{5}$, $10$ with $\bar{10}$ in the $\rm{SU}(5)$ 
or $16$ with $\bar{16}$ in the $\rm{SO}(10)$ respectively \cite{Zheng1, Zheng2}.
A realistic embedding should satisfy the following constraints.
\begin{itemize}
\item The vev of Higgs field $\rho$ responsible for the VL fermion masses should be of order TeV scale.
\item The VL fermions are prevented from directly coupling to the Higgs fields triggering high-scale gauge symmetry breaking.
\end{itemize}
Violating the first constraint is likely to occur because
either $\rho=\{1,24, 75\}$ or $\rho=\{1, 45,210\}$
may directly couple to $S$,$S'$ $Z$, $Z'$ in the case of $\rm{SU}(5)$ or $X$ and $Y$ in the case of $\rm{SO}(10)$, respectively,
which tends to yield $\langle\rho\rangle$ of GUT scale.
In contrast, $\langle\rho\rangle$ of order TeV scale can be only realized by the effective operator such as
\begin{eqnarray}{\label{vvev1}}
W_{\text{eff}}(\rho)=\frac{M_{\rho}}{2}\rho^{2}+\frac{\rho\cdot A\cdot B\cdots}{M^{n}_{\text{U}}}
\end{eqnarray}
where $A$, $B$, $\cdots$ refer to $H(5)$, $\bar{H}(\bar{5})$, $S$, $S'$, $Z$, $Z'$ in the $\rm{SU}(5)$,
or $H(126)$, $\bar{H}(\bar{126})$, $X$ and $Y$ in the $\rm{SO}(10)$, respectively, 
with $M_{U}$ denoting the GUT scale.

\subsection{\rm{SU}(5)}
For the VL fermions of $5(\Sigma)$ and $\bar{5}(\bar{\Sigma})$, 
$W_Y$ in Eq.(\ref{Yukawa5}) is extended by,
\begin{eqnarray}{\label{VMSSM5}}
\delta W_{Y}=\rho\bar{\Sigma}(\bar{5})\Sigma(5)+\text{H.c},
\end{eqnarray}
where $\rho=\{1,24\}$ of $\rm{SU}(5)$. 
The reason for adding $\rho$ is that either singlet vev $S$ or $S'$ in the Sec.IIA is too large to provide a VL mass of order TeV.

In this case, the unsafe superpotential at least includes
\begin{eqnarray}{\label{VMSSMu}}
W_{\text{unsafe}}\supset S\Sigma\bar{\Sigma},
\end{eqnarray}
which can be excluded by imposing the $Z_2$ parity assignments as shown in Table.\ref{VMSSMc1}.
$W_{\text{unsafe}}$ in Eq.(\ref{VMSSMu}) can also contain the following terms depending on $\rho$, 
\begin{eqnarray}{\label{VMSSMu2}}
W_{\text{unsafe}}=
 \left\{
\begin{array}{lcl}
\rho Z^{2}+\rho Z'^{2}+S\rho Z'+\rho^{2}S, ~\rho=24\\
\rho Z^{2}+\rho Z'^{2}+\rho^{2}S+\rho S^{2}, ~~~\rho=1\\
\end{array} \right. 
\end{eqnarray}

\begin{table}
\tiny{
\begin{center}
\begin{tabular}{|c|c|c|c|c|c||c|c|c|c||c|c|c|c|c|c|c|c|}
\hline
 Field  & $N_{R}$&\ $\psi$ & \ $\Phi$ & $\Sigma$ & $\bar{\Sigma}$ & \  $5$ & \ $\bar{5}$ & \ $\bar{45}$ & $\rho$ & \ $1(S)$ & $1(S')$ &\ $75(Z)$ & $75(Z')$  & $45$ & \ $50$ &  $\bar{50}$   \\
  \hline
$Z_{2}$ &+ & + &  - & + & - & + & -  & - & -  & + & + &  +  & - & -& - & - \\
  \hline
\end{tabular}
\caption{$Z_{2}$ parity assignments for embedding NMSSM into $\rm{SU}(5)$,
which is consistent with the superpotential in Eq.(\ref{VMSSM5}).
Yet, it is unable to exclude all the unsafe structures in Eq.(\ref{VMSSMu2}). }
\label{VMSSMc1}
\end{center}}
\end{table}

Besides the unsafe operators in Eq.(\ref{VMSSMu2}), 
there are also no suitable Feynman graphs to generate the desired effective operator with correct mass order in Eq.(\ref{vvev1}).
In principle, the form of effective operator in  Eq.(\ref{vvev1}) can be divided as follows:
\begin{itemize}
\item It is of type $\rho AB\cdots /M^{2}_{\text{U}}$, with $A$ and $B$ referring to $S$, $Z$ or $Z'$.
In this situation, one obtains $\langle\rho\rangle \sim$ GUT scale given all of $A$, $B$, $\cdots$ are of order GUT scale.
\item It looks like $\rho AB\cdots H(5)/M_{\text{U}}$ or $\rho AB\cdots\bar{H}(\bar{5})/M_{\text{U}}$, 
which contain only a doublet vev.
Such operator contributes to $\langle\rho\rangle\sim$ TeV for $M_{\rho}\sim$ GUT scale.
Unfortunately, there is no suitable intermediator.
\item It is of form $\rho AB\cdots H(5) \bar{H}(\bar{5})\cdots/M^{2}_{\text{U}}$ or $\rho AB\cdots \bar{H}(\bar{5})H(45)\cdots/M^{2}_{\text{U}}$, 
where there are at least two doublet vevs.
In this case, $\langle\rho\rangle$ is less than $\mathcal{O}(1)$ eV.
\end{itemize}
In summary, since there is no appropriate vev scale, the embedding of VMSSM into $\rm{SU}(5)$ is not viable.
Similar result can be found for VL $10$ and $\bar{10}$, where $\rho=\{1,24, 75\}$ of $\rm{SU}(5)$. 
\\
\\

\subsection{\rm{SO}(10)}
As shown in \cite{Zheng2}, 
the non-minimal extension through the $16$-dimensional VL fermions remains consistent with the $\rm{SO}(10)$ unification.
In this model, $W_Y$ is modified by,
\begin{eqnarray}{\label{VMSSM10}}
\delta W_{Y}=\rho \Delta(16)\bar{\Delta}(\bar{16}),
\end{eqnarray}
where $\rho=\{1, 45, 210 \}$ of $\rm{SO}(10)$. 
Similar to Sec.IVA,
the main challenge to the embedding is the generation of singlet vev $\langle\rho\rangle\sim$ TeV scale for a $\rho$ mass of GUT scale.
What differs from the discussions in Sec.IVA is the existence of the intermediate mass scales 
$\upsilon^{126}_{s}$ and $\upsilon^{\bar{126}}_{s}$ in the $\rm{SO}(10)$,
which is critical to solve the problem.

\begin{figure}
\centering
\includegraphics[width=5cm,height=4cm]{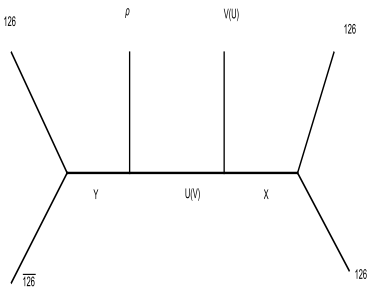}
\centering
 \caption{Super Feynman graph for the generation of higher-dimensional effective operator in the case of $\rho=210$.}
\label{vmssm}
\end{figure}

We firstly consider $\rho=210$.
We add Higgs fields $54(V)$ and $54(U)$ to $W_{\text{SB}}$, 
with the $Z_{2}$ parity assignments as shown in Table.\ref{VMSSMc2}.
The $Z_2$ parity excludes the unsafe operator
\begin{eqnarray}{\label{unsafev}}
W_{\text{unsafe}}=Y\Delta(16)\bar{\Delta}(\bar{16}),
\end{eqnarray}
and simultaneously allows Yukawa interactions
\begin{eqnarray}{\label{SSBM10V1}}
\delta W_{\text{SB}}&=&\frac{M_{\rho}}{2}\rho^{2}+\frac{M_{V}}{2}V^{2}+\frac{M_{U}}{2}U^{2}\nonumber\\
&+&\lambda'_{10}\rho UY+\lambda''_{10}\rho VY \nonumber\\
&+&\lambda'_{8}UVX+\lambda''_{8}U^{2}X+\lambda'''_{8}V^{2}X.
\end{eqnarray}
In terms of Eq.(\ref{SSBM10V1}) and Eq.(\ref{SSBM10}) the effective operator for $\rho$ is given by,
\begin{eqnarray}{\label{operator}}
W_{\text{eff}}&\simeq& \frac{M_{\rho}}{2}\rho^{2}+ \rho(\lambda'_{10}U+\lambda''_{10}V)\left[\frac{H(126)\bar{H}(\bar{126})}{M_{Y}}+\cdots\right]\nonumber\\
&+&\rho\frac{\lambda'_{10} U H^{3}(126)\bar{H}(\bar{126})}{M_{Y}M_{X}M_{V}}+\rho\frac{\lambda''_{10} V H^{3}(126)\bar{H}(\bar{126})}{M_{Y}M_{X}M_{U}}\nonumber\\ &+& \mathcal{O}(\rho^{3})
\end{eqnarray}
For calculating an effective superpotential in the infrared region from those in the ultraviolet region,
integrating out heavy chiral superfields in the Feynman graphs is equivalent to solving the nonlinear equations of $F$ terms related to these heavy chiral superfileds.
The leading-order terms with coefficient $\lambda'_{10}U+\lambda''_{10}V=\delta F_{Y}$ in Eq.(\ref{operator}) are obtained after integrating out superfield $Y$ for $\lambda_1$ and $\lambda_{10}$ less than unity.
Similarly, the next-leading order operators therein are induced by further integrating out $X$,
referring to which the Feynman graph is shown in Fig.\ref{vmssm}.
Note, we have used the mass term in Eq.(\ref{operator}) representing those quadratic terms,
and neglected the higher-order terms.

Apart from the $F$-term contributions in Eq.(\ref{operator}),
the potential for singlet component in $\rho$ also contains soft SUSY-breaking terms such as $A_{\rho}\rho^{3}+\text{H.c}$.
The scale of $A_{\rho}$ depends on the details of SUSY breaking.
It can be neglected in gauge mediation, but be of order $\sim 1$ TeV in the other scenarios. 
In the later case, it is easily to verify that the $F$-term contribution still dominates over those soft terms.
In what follows, we simply ignore those soft terms for the estimate of $\langle\rho\rangle$. 
 
The leading-order contribution in Eq.(\ref{operator}) has to be eliminated if one wants to obtain singlet vev $\langle\rho\rangle\sim$ TeV scale.
Alternatively, the coefficient $\delta F_{Y}$ must be suppressed without much fine tuning.
There is a dynamical realization for this purpose.
From Eq.(\ref{SSBM10V1}), 
one finds that the SUSY vacuum described by the vevs in Eq.(\ref{vevs3}) remains 
only if the following constraints 
\begin{eqnarray}{\label{vevs4}}
0&=&\lambda'_{8}UV+\lambda''_{8}U^{2}+\lambda'''_{8}V^{2},\nonumber\\
0&=&\lambda'_{10}U+\lambda''_{10}V,\nonumber\\
0&=&M_{V}V+\frac{X}{2\sqrt{15}}\left(\lambda'_{8}U+2\lambda'''_{8}V\right),\nonumber\\
0&=&M_{U}U+\frac{X}{2\sqrt{15}}\left(\lambda'_{8}V+2\lambda''_{8}U\right),
\end{eqnarray}
are satisfied. 
Given GUT-scale mass parameters $M_U$, $M_V$, $\langle U\rangle$ and $\langle V\rangle$,
there are indeed rational solutions to Eq.(\ref{vevs4}),
under which the leading-order operator in Eq.(\ref{operator}) vanishes due to $\delta F_{Y}=0$ (the second formula in Eq.(\ref{vevs4})).
Therefore, the next leading-order contribution in Eq.(\ref{operator}) dominates 
the effective superpotential for $\rho $ below GUT scale as long as $M_{U}\neq M_{V}$,
\begin{eqnarray}{\label{vvev3}}
W_{\text{eff}}\sim M_{\rho}\rho^{2}+\rho \frac{V H^{3}(126)\bar{H}(\bar{126})}{M^{3}_{U}},
\end{eqnarray}
which contributes to a nonzero vev: 
\begin{eqnarray}{\label{rho}}
\langle\rho\rangle \sim \frac{\langle V\rangle}{M_{\rho}} \frac{\left(\upsilon^{126}_{s}\right)^{4}}{M^{3}_{U}}.
\end{eqnarray}
Given singelt vev $\upsilon^{126}_{s}\simeq \upsilon^{\bar{126}}_{s}\sim 10^{13}$ GeV fixed by fit to SM flavor masses \cite{1805.10631}, 
and $M_{\rho}\sim M_{U}\sim \langle V\rangle \sim 10^{16}$ GeV,
we have $\langle\rho\rangle \sim 1-10$ TeV.

\begin{table}
\begin{center}
\begin{tabular}{|c|c|c|c||c|c|c||c|c|c|c|c|c|}
\hline
 Field & \ $\phi$&\ $\Delta$ & \ $\bar{\Delta}$   & \  $10$ & $\bar{126}$ & $\rho$ & \ $X$ & \ $126$&\ $54(V)$ & $54(U)$ & $Y$ \\
\hline
$Z_{2}$ & +  &  + & - &  +  &  + &  - & +  & +  & - & - & + \\
  \hline
\end{tabular}
\caption{$Z_{2}$ parity assignments in the VMSSM model with $\rho=210$,
which are consistent with the superpotentials in Eq.(\ref{Yukawa10}),Eq.(\ref{SSBM10}), Eq.(\ref{VMSSM10}) and Eq.(\ref{SSBM10V1}),
and simultaneously avoid the unsafe operator in Eq.(\ref{unsafev}). }
\label{VMSSMc2}
\end{center}
\end{table}

The analysis for $\rho=210$ can shed light on other cases such as $\rho=\{1,45\}$.
For $\rho=1$ we can naively choose  $Z_2$ odd fields $U=V=210$ in Fig.\ref{vmssm}.
However, an unsafe operator $V(U)\Delta(16)\bar{\Delta}(\bar{16})$ appears again.
For $\rho=45$, one may choose $U=45$ and $V=45$ or $54$, 
which is unfavored by an operator similar to $\rho=1$.

\section{Conclusion}
According to the observed Higgs mass at the LHC and the dark matter direct detection limits,
the conventional MSSM - the simplest natural SUSY that is consistent with unification - is  under more pressure than ever.
Such stress can be greatly relaxed in the extended MSSM models such as NMSSM and VMSSM which retain the unification and are still simple.
In this paper, following the assumptions that $W_Y$ is fixed by the SM matter content and its TeV-scale extension, 
and they receive their masses from $W_{\text{SB}}$ through Higgs mechanism,
we have studied the embeddings of these three models into SUSY GUTs.

First of all, we discussed the MSSM, where the realistic $\rm{SU}(5)$ and $\rm{SO}(10)$ realizations 
serve as benchmark solutions to the SM flavor issue and neutrino masses.
Then, we utilize the benchmark MSSM as guidance to the embedding of NMSSM and VMSSM.
We have found that the embedding of NMSSM is not viable 
due to a large mount of mixings between the singlet $N$ and the Higgs fields responsible for the GUT symmetry breaking.
But the problem can be evaded in the VMSSM, 
because the Higgs field $\rho$ which provides $16$-dimensional VL mass of order TeV scale 
can avoid the same problems the singlet $N$ encounters due to the intermediate mass scale in the $126$-dimensional Higgs.

{\it Acknowledgments}.
The author thanks the anonymous referee for useful suggestions.
This work is supported by the National Natural Science Foundation of China under Grant No.11775039 
and the Fundamental Research Funds for the Central Universities with project No. cqu2017hbrc1B05 at Chongqing University.

\end{document}